# Efficient Read-Port-Count Reduction Schemes for the Centralized Physical Register File in a Superscalar Microprocessor

Denis Los

*Abstract—* The physical register file supports increasing the execution width and depth of a superscalar microprocessor to exploit more instruction-level parallelism. The efficient design of the physical register file is critical since its resources, such as the number of read and write ports, have a significant impact on CPU power consumption. Reducing the number of ports to the physical register file is a well-known direction for optimization. For port-count reduction schemes, balancing the trade-off between the scheme's complexity and performance is crucial. In our work, we introduce a high-level analysis method to estimate the complexity of the schemes during microarchitectural design. Moreover, we explore the structure of different port-count reduction schemes and introduce a practical approach to constructing low-complexity read-port-count reduction schemes for the centralized integer physical register file. We show that the read-port-count reduction schemes designed with this approach can reduce the number of read ports by a factor of two (from 17 to 8 read ports) with the Geomean performance degradation of only 0.1% IPC across the SPECrate CPU 2017 Integer workloads.

## I. Introduction

Modern superscalar microprocessors with out-of-order execution tend to improve performance through increasing instruction-level parallelism. Hardware structures such as the physical register file (PRF) are the cornerstones to exploiting more instruction-level parallelism in an out-of-order machine. Increasing the execution depth requires more entries in the physical register file in order to support register renaming and store the values. Increasing the execution width of a superscalar machine requires the addition of more read and write ports to the physical register file. This leads to high pressure on the physical register file and challenges in designing an energy-efficient and area-efficient processor. The number of read and write ports significantly impacts the required area and power consumption for the register file, since there is usually a quadratic relation between the area and the number of register file ports [1]-[3].

Due to the high importance of the topic, there have been many works dedicated to the design, organization, and optimization of register files in general-purpose processors [4]. In this work, similar to [5], we focus on reducing the number of ports to the centralized PRF of a general-purpose superscalar processor in contrast to exploring multi-bank designs. We primarily target the reduction of read ports to the register file, since their number is usually bigger.

Instructions are usually sent to the execution to dedicated functional units through the corresponding issue ports. If some of the read ports to the PRF are removed, several of the remaining read ports will be shared between different issue ports. Naturally, in the sharing read-port-count reduction scheme, different issue ports cannot use the same read port simultaneously. If there is a conflict and multiple issue ports need the same shared read port to the PRF to get the source operand values, sending some instructions to the execution will be canceled this cycle and postponed. Usually, sending these canceled instructions to the functional execution units will be retried the next processor cycle. However, these cancellations result in performance degradation from the read-port-count reduction schemes.

There are several key reasons why conflicts for read ports to the PRF do not happen all the time and performance degradation from introducing read-port-count reduction schemes can be mitigated. First, due to stalls in a processor pipeline caused by such events as branch mispredictions, cache misses, and structural hazards, there might be no ready instructions to be sent through some of the issues ports in some processor cycles. It opens the opportunity for other issue ports to freely use the non-utilized read ports to the PRF in these cycles. Our analysis shows that in a superscalar processor with single-thread out-of-order execution, the average utilization of the execution width might be as low as 50% on critical workloads. Second, instructions do not need to read all of their source operands from the register file all the time.

There are multiple reasons why instructions might not need to read all of their operands from the register file. For example, an instruction reading an immediate value will usually either propagate it directly from the front-end decoding logic or take it from the separate data array that stores immediate values. Moreover, if an instruction takes a register value from a bypass, it does not need to use a read port to the PRF. However, our analysis shows that only 40% of integer operands get their values from the bypass network. Furthermore, in some architectures, such as x86-64, some

Denis Los – Moscow Institute of Physics and Technology (9 Institutskiy per., Dolgoprudny, Moscow Region, 141700, Russian Federation)
ORCID: https://orcid.org/0009-0009-4500-8106
email: los.da@phystech.edu

instructions do not need all of their operands. Memory access instructions in x86-64 architecture usually take two register operands, the base and index operands. The linear effective address for memory access is computed using a base value, an index value with a scale, and an immediate displacement value. However, for example, if the index operand is not provided, the address will be computed using only the base register. Only the value of the base register will be read from the PRF using a read port. Therefore, reducing PRF read ports corresponding to issues ports for memory access instructions is the natural and common approach.

There is a trade-off between area reduction and performance degradation in the read-port-count reduction schemes. However, the latency of the port arbitration logic in the read-port-count reduction schemes for the PRF is also an extremely important and frequently overlooked parameter. The more different issue ports compete for the same read port to the PRF in a read-port-count reduction scheme, the more complex the logic to detect conflicts and assign the shared read ports becomes. If the latency of the critical path of the logical scheme implementing the read-port-count reduction optimization becomes too high, it might require the addition of new pipeline stages and updates throughout the dynamic scheduling logic. Naturally, it will also lead to higher performance degradation from the read-port-count reduction. Furthermore, the issue logic is on the critical path of the whole processor pipeline, hence, adding the extra complex logical schemes will result in higher power consumption.

In this work, we make the following main contributions.
1) We introduce a practical high-level analysis method to estimate the complexity of read-port-count reduction schemes during the microarchitectural design.
2) We introduce a subset of read-port-count reduction schemes called *uniform symmetric schemes* and show the benefits of utilizing such schemes
3) We propose a set of practical rules to construct efficient uniform symmetric read-port-count reduction schemes that minimize the number of conflicts for read ports.

II. RELATED WORK

As we mentioned, there is a large scope of work dedicated to designing efficient CPU register files. A great survey on the relevant techniques has been conducted in [4]. In this paper, we focus on read-port-count reduction schemes for the centralized monolithic integer physical register files in CPUs. As highlighted in [4], the other techniques to reduce the pressure on the PRF and optimize its resources include introducing hierarchical designs of register files [6]-[8], using multi-banked register file organizations [8]-[12], as well as the register cache [1], [13]-[15]. Furthermore, compiler-assisted optimizations, utilization of data locality, improvements to the bypass network, optimizations to the register allocation and release, and leveraging narrow-width values are the order common approaches [4].

In [5], approaches to reducing the number of read ports in the centralized integer physical register files are explored. Limitations of the brute-force approaches to read-port-count reduction are highlighted. Hence, several optimizations such as leveraging the narrow-width values, introducing clever priorities in the port arbitration logic, and splitting the instructions between the operands to improve the cooperation with a data forwarding network are proposed. In our work, we introduce several methods to construct brute-force schemes minimizing their complexity. Our approaches could be used with the optimizations introduced in [5]. In [16], read-port-count reduction techniques are explored for the register file in a VLIW processor. Interconnection topologies for shared-port register files are classified into complete, direct, and partial interconnection schemes. An algorithm is introduced to construct efficient direct interconnection schemes. The approach to the constructing of the efficient direct interconnection schemes in [16] is very similar to ours. However, we use our approach to construct both direct and partial interconnection schemes.

III. METHODOLOGY

For all experiments in this paper, we use a cycle-accurate x86-64 performance simulator. We model a superscalar CPU core with single-thread out-of-order execution. The modeled CPU core has 9 functional execution units that perform operations on integer operands: 4 arithmetic logic units (ALU), 2 load execution functional units with address generation units (AGU), 2 functional units with AGUs for store-address (STA) operations, and 1 functional store-data unit to perform store-data (STD) operations. The number of read ports to the integer PRF in the baseline scheme with no reductions is 17: 8 read ports for the ALUs, 8 read ports for the all AGUs, and 1 read port for the store-data functional unit. The number of write ports to the integer PRF is 10: 8 write ports for the ALUs and 2 write ports for the load execution units.

The CPU model has the integer physical register file with 180 entries, the reorder buffer with 224 entries, and the reservation station with 97 entries. The sizes of the load and store queues are 56 and 72 entries, respectively. The CPU model has 32 KiB 8-way L1 instruction and data caches and 256 KiB 4-way L2 cache.

All of the experiments are conducted on the workloads from the SPECrate CPU 2017 Integer benchmark suite [17] compiled for the Linux operating system using GCC compiler.

IV. CONSTRUCTING READ-PORT-COUNT REDUCTION SCHEMES

In this work, we explore the read-port-count reduction schemes that reduce the number of read ports to the integer PRF from 17 to 8. We remove 9 read ports assigned to the functional units that execute load and store operations. Hence, 4 ALUs and 5 functional units that execute load and store instructions start to share the remaining 8 read ports to the PRF.

If several functional units are competing for the same read port, multiple approaches can be used to determine which functional unit should utilize the port. To reduce the complexity of the read-port-count reduction schemes, in this paper, we consider the following approach. We statically assign priorities to each of the functional execution units. ALUs receive the highest priority level. We do not share read ports between different ALUs, hence, if an ALU requires a read port, it will use its dedicated port. Load and store execution units will be able to use a read port, only if it is not requested by an ALU. We assign the second priority level to the load execution units. The third and fourth priority levels are assigned to the store-address and store-data execution units, respectively. Furthermore, we assign internal priority levels to the execution units of the same type. For example, if an ALU does not require a read port, which load execution unit will use the port will be determined based on the statically assigned priority levels.

Similar to [16], we introduce a matrix as a convenient way to represent real-port-count reduction schemes and describe the interconnections in the port arbitration logic. An example of such a matrix is provided in Fig. 1. The indexes of the functional execution units (0-3) and the corresponding indexes of the remaining read ports (0-7) are written at the top. The indexes of the load and store execution units and their operands are written on the left. $s1$ and $s2$ denote the first and the second operands, respectively. Unit 8 here corresponds to the store-data function unit that can read only one operand. Units 4 and 5 are the load execution units, while units 6 and 7 are the store-address functional units. The 1s in the matrix represent connections between the remaining read ports to the integer PRF and the operands of the load and store execution units with reduced read ports. The 0s indicate the absence of such connections.

| C | | 0 | | 1 | | 2 | | 3 | |
|---|---|---|---|---|---|---|---|---|---|
| | | 0 | 1 | 2 | 3 | 4 | 5 | 6 | 7 |
| 4 | s1 | 0 | 1 | 0 | 0 | 1 | 0 | 0 | 0 |
| | s2 | 0 | 0 | 1 | 0 | 0 | 0 | 0 | 1 |
| 5 | s1 | 0 | 0 | 0 | 1 | 0 | 0 | 1 | 0 |
| | s2 | 1 | 0 | 0 | 0 | 0 | 1 | 0 | 0 |
| 6 | s1 | 0 | 0 | 1 | 0 | 0 | 0 | 0 | 1 |
| | s2 | 0 | 1 | 0 | 0 | 1 | 0 | 0 | 0 |
| 7 | s1 | 1 | 0 | 0 | 0 | 0 | 1 | 0 | 0 |
| | s2 | 0 | 0 | 0 | 1 | 0 | 0 | 1 | 0 |
| 8 | s2 | 1 | 0 | 0 | 0 | 0 | 1 | 0 | 0 |

Figure 1. Example of the matrix representing the read-port-count reduction scheme

For example, for the scheme in Fig. 1, the first operand (the base operand) of the load instruction on the 4th issue port and the second operand (the index operand) of a STA instruction on the 6th issue port can be read from the read ports with indexes 1 and 4. However, it is possible only if these read ports are not requested by instructions executing on ALUs on the 0th and 2nd issue ports. For the base operand of the load instruction on the 4th issue port, the value will be read from read port 1 if the read port is not used by the ALU and the load instruction needs the base value to be read. The requested base operand value will be read from read port 4 if the port is free and read port 1 is occupied. For the requested index value of the STA instruction on the 6th issue port, the value will be read from port 1 only when the read port is not used by the ALU and the base operand of the load instruction on the 4th issue port is not requested. The requested index value on the 6th issue port will be read from port 4 only when, first, the arithmetic instruction executing on the 2nd issue port does not read from port 4, and, second, either read port 1 is not used by the 0th issue port and the base operand on 4th issue port is requested or read port 1 is used by the 0th issue port and the base operand on 4th issue port is not requested.

For each of the remaining read ports, port arbitration logic needs to determine which functional unit will use it to get the operand value this cycle and whether any instructions need to be canceled. For each read port, let us define a one-hot vector $s[0..16]$ showing which operand will be read using this read port. In this one-hot vector, indices from 0 to 7 correspond to the operands of the ALUs, and indices from 8 to 16 correspond to the operands of the memory access functional units. We define a matrix $S[0..7][0..16]$ as the combination of one-hot vectors for each read port. Let us denote as $a[0..7]$ the vector showing which operands are requested by ALUs. We denote as $k[0..8]$ the vector showing which operands are requested by the memory access functional units. Using vectors $a[0..7]$ and $k[0..8]$, the values in the matrix $S[0..7][0..16]$ could be computed for each processor cycle.

## V. Estimating Complexity of the Read-Port-Count Reduction Schemes

It is necessary to estimate the complexity of the port arbitration logic based on the critical path. The real critical path of the functional logic in the hardware depends on many parameters, such as the types of transistors used, the processor frequency, the platform for implementing the hardware logic, the accuracy of the circuit synthesis algorithms, and many others. In this regard, in practice, when developing the processor microarchitecture, it is not possible to estimate the real critical path of the logical circuit, since most of these parameters become known only at the stage of hardware logic design when the main work on developing the processor microarchitecture has already been completed.

Thus, it is necessary to introduce a computable for port arbitration logic of read-port-count reduction schemes, allowing us to estimate the length of the critical path during microarchitectural design and simulation stages. The length of the critical path can be determined by estimating the complexity of logical expressions used to calculate the elements of the matrix $S[0..7][0..16]$. We use an element in the matrix $S$ with the longest critical path as the estimation for the critical path of the whole port arbitration logic.

Elementary logic elements used at the stage of microprocessor design can vary in the number of supported input signals depending on the type of transistors used and their size. Specific sets of elementary logic elements and their characteristics, as noted earlier, become known, usually, only at the stage of hardware design. In addition, the synthesis of logic circuits with arbitrary non-uniform elements is a complex computational problem and is solved using large computing clusters at the stage of circuit design. Thus, when constructing logical functions for estimating the critical path, we will use elementary logic elements: a negation element, a dual-channel AND element, and a dual-channel OR element. We consider the lengths of the critical paths of the elementary conjunction and disjunction elements to be the same and equal to one-time quantum. To simplify the calculations, we will neglect the length of the critical path of the negation element compared to the lengths of the critical paths of AND and OR elements.

To determine the length of the critical path of a logical function, it is not enough to simply consider an arbitrary implementation of the function using negation, conjunction, and disjunction elements, but it is necessary to use a minimal representation with the shortest critical path length. It is known that the minimal representations for a logical function implemented using negation, conjunction, and disjunction elements are the Minimum Conjunctive Normal Form (MCNF) and the Minimum Disjunctive Normal Form (MDNF), represented using these elements as a binary tree. The depths of the constructed trees for MCNF and MDNF will be the minimal critical paths. By selecting the minimal depth from the trees constructed using MCNF and MDNF, the desired critical path length for the logical function will be obtained.

There are many algorithms for finding the MCNF and MDNF from the constructed logical functions for the elements of the matrix S, differing in their running time and minimization accuracy. In the case of a small number of input elements, it is possible to minimize the functions using methods that allow finding an exact solution, for example, the Quine-McCluskey method, which is easily implemented using different programming languages. However, the running time of the Quine-McCluskey algorithm increases exponentially with an increase in input data. Thus, for logical circuits with a large number of input parameters (for microprocessors with a large number of functional execution units), it is necessary to use heuristic minimization methods, such as the Espresso algorithm [18].

As a result, the algorithm for estimating the length of the critical path for read-port-count reduction schemes in a physical register file could be described as follows.

1. For each calculated element of the matrix S, construct a logical function.

2. Using one of the minimization algorithms, construct the MCNF and MDNF for elements from S.

3. Represent the MCNF and MDNF as binary trees and determine the depth of each of them.

4. Select the minimum depth from the MCNF and MDNF representations.

With the help of this proposed algorithm, it is possible to estimate the complexity of the port arbitration control logic of read-port-count reduction schemes and compare different schemes with each other. However, by itself, this metric does not give an absolute threshold value, the excess of which will mean that the port arbitration control logic will not be able to perform calculations within one processor cycle.

Such threshold value can be estimated if we find a logical function with the boundary length of the critical path under the conditions of the existing technological process. For example, we can find the logical function that needs approximately one cycle to be fully computed in the hardware of a current processor. Having found this logical function, we can estimate the length of the critical path for it using the proposed algorithm. The estimated critical path length for the analyzed logical functions can be used as the threshold value.

## VI. EFFICIENT READ-PORT-COUNT REDUCTION SCHEMES

Read-port-count reduction schemes can lead to significant performance degradation caused by a large number of conflicts for the read ports.

According to our analysis, the performance degradation can reach up to -4% IPC (Instructions per Cycle) on average (and up to -30% IPC on individual workloads) with poor organization of the schemes.

The complexity of the read port reduction schemes depends heavily on the arrangement of the functional connections between the memory access operands and the read ports to the physical register file. For example, in Fig. 2 and Fig. 3 we show the schemes for reducing the read ports from 17 to 8 ports.

| C | | 0 | | 1 | | 2 | | 3 | |
|---|---|---|---|---|---|---|---|---|---|
| | | 0 | 1 | 2 | 3 | 4 | 5 | 6 | 7 |
| 4 | s1 | 1 | 0 | 0 | 0 | 0 | 0 | 0 | 0 |
| | s2 | 0 | 1 | 0 | 1 | 0 | 0 | 0 | 0 |
| 5 | s1 | 0 | 0 | 1 | 0 | 0 | 0 | 0 | 0 |
| | s2 | 0 | 1 | 0 | 1 | 0 | 0 | 0 | 0 |
| 6 | s1 | 0 | 0 | 0 | 0 | 1 | 0 | 0 | 0 |
| | s2 | 0 | 0 | 0 | 0 | 0 | 1 | 0 | 0 |
| 7 | s1 | 0 | 0 | 0 | 0 | 0 | 0 | 1 | 0 |
| | s2 | 0 | 0 | 0 | 0 | 0 | 0 | 0 | 1 |
| 8 | s2 | 0 | 0 | 0 | 0 | 0 | 0 | 1 | 0 |

Figure 2. Example of the read-port-count reduction scheme with a complexity of 3 time units

| C | | 0 | | 1 | | 2 | | 3 | |
|---|---|---|---|---|---|---|---|---|---|
| | | 0 | 1 | 2 | 3 | 4 | 5 | 6 | 7 |
| 4 | s1 | 1 | 0 | 0 | 0 | 0 | 0 | 0 | 0 |
| | s2 | 0 | 1 | 0 | 0 | 0 | 0 | 0 | 0 |
| 5 | s1 | 0 | 0 | 1 | 0 | 0 | 0 | 0 | 0 |
| | s2 | 0 | 1 | 0 | 1 | 0 | 0 | 0 | 0 |
| 6 | s1 | 0 | 0 | 0 | 0 | 1 | 0 | 0 | 0 |
| | s2 | 0 | 0 | 0 | 0 | 0 | 1 | 0 | 0 |
| 7 | s1 | 0 | 0 | 0 | 1 | 0 | 0 | 1 | 0 |
| | s2 | 0 | 0 | 0 | 0 | 0 | 0 | 0 | 1 |
| 8 | s2 | 0 | 0 | 0 | 0 | 0 | 0 | 1 | 0 |

Figure 3. Example of the read-port-count reduction scheme with a complexity of 4 time units

Both schemes have the same total number of functional connections and differ only in the location of one functional connection. However, the estimated critical path length for the scheme in Fig. 2 is 3 time units, while for the scheme in Fig. 3, it is 4 time units.

This difference in the values of the critical path lengths is explained by the difference in the number of dependent parameters entering different elements of the matrix S. In cases where the parameters from the vectors a and k are uniformly included in the logical expressions for different elements of the matrix S, the estimate of the critical path length is usually not high and does not exceed the threshold values. However, in cases where the uniform distribution is violated and a large number of parameters are included in one or more logical expressions calculated for S, the critical path length becomes large and can exceed the threshold value for many functional connections. The rows of the representation matrix of the read-port-count reduction scheme are vectors that reflect the functional connections between the operands used in memory access functional units and the read ports from the physical register file. Let us consider the set of unique rows of the representation matrix. Each vector entering the set of unique rows will be called a *mask* of the read-port-count reduction scheme. The number of masks for a particular read-port-count reduction scheme is the total number of unique vectors entering the set of rows of the representation matrix. For the scheme shown in Fig. 2, the number of masks is 7, and for the scheme shown in Fig. 3, the number of masks is 9.

In addition, let us consider the mask configurations for the two schemes shown in Fig. 2 and Fig. 3. It can be seen that the masks (unique rows of the representation matrix) for the scheme in Fig. 2 do not intersect with each other, and their union gives a single row, which means that each of the available read ports has at least one functional connection to the operands of the memory access units. The masks for the scheme in Fig. 3 have pairwise intersections, shown in Fig. 4, although they completely cover the set of read ports with functional connections.

| 0 | 1 | 0 | 0 | 0 | 0 | 0 | 0 |
|---|---|---|---|---|---|---|---|
| 0 | 1 | 0 | 1 | 0 | 0 | 0 | 0 |
| 0 | 0 | 0 | 1 | 0 | 0 | 1 | 0 |
| 0 | 0 | 0 | 0 | 0 | 0 | 1 | 0 |

Figure 4. Example of the intersecting masks

The intersection between masks in read-port-count reduction schemes forms dependencies between different calculated elements of the matrix S. Although the logical expressions for the elements of the matrix S are calculated in parallel and do not directly lead to an increase in the length of the critical path of the logical expression, they lead to the dependence of the matrix element function on a larger number of parameters from the vectors a and k, thereby complicating the control logic of the scheme.

The intersection between masks affects not only the complexity of the port arbitration logic but also the system performance when using read-port-count reduction schemes. This is explained by the fact that in the case of the intersection, some of the read ports are loaded more heavily, leading to an increase in the number of conflicts for the operands used in functional units that have a low priority on this port, while some of the read ports are not fully utilized. Also, to improve the performance when using port reduction techniques, the set of masks should cover the largest part of the set of read ports, to avoid a situation with a similar unbalanced load on the read ports, leading to an increase in the number of conflicts for read ports from an integer physical register file. Thus, to build schemes that are efficient in terms of the complexity of the port arbitration logic and system performance for a given total number of functional connections between the operands used in the memory access functional units and the read ports, it is necessary to use masks that satisfy the following conditions:

1. The union of masks yields a unit row vector of the size equal to the number of non-reduced available read ports
2. The masks do not intersect with each other

We will call read-port-count reduction schemes the masks of which satisfy the above conditions *symmetric*. Note that in general, this does not mean geometric symmetry in the representation matrix of read-port-count reduction schemes.

In symmetric read-port reduction schemes, different masks may correspond to different numbers of functional connections, thus leading to different numbers of read conflicts for operands with the same utilization. Hence, for the workloads in which some of the operands used in the execution units with masks with the smallest number of functional connections are used more than others, the performance drop will be greater. Therefore, to maintain the same average performance on different workloads, symmetric schemes should be used, the masks of which correspond to the same number of functional connections. We will call such symmetric schemes *uniform symmetric*.

The number of functional connections in masks of uniform symmetric read-port-count reduction schemes cannot be arbitrary. Since the mask vectors should not intersect with each other and must completely cover the set of available read ports, then, consequently, the number of functional connections in masks of uniform symmetric schemes must be a divisor of the number of non-reduced read ports. For the considered examples, when the number of read ports is reduced from 17 to 8, masks of uniform symmetric schemes can have 1, 2, or 4 functional connections. Naturally, masks with 8 functional connections are not practically significant.

All uniform symmetric schemes with a given number of connections in masks and with a fixed number of operands in memory access functional units for which the read ports have been reduced will have the same estimate of the length of the critical path of the port arbitration logic. This is the result of the constraints imposed on uniform symmetric schemes. Examples of uniform symmetric read-port-count reduction schemes with the number of connections in masks equal to 1, 2, and 4 are shown in Fig. 5, Fig. 6, and Fig. 7, respectively. The corresponding estimates of the lengths of the critical paths are equal to 2, 5, and 9 time units.

However, using arbitrary uniform symmetric schemes is not enough to obtain an acceptable performance degradation with read-port-count reduction schemes. The scheme shown in Fig. 5 leads to a geomean performance drop of -3.2% IPC.

In uniform symmetric schemes with a fixed number of connections, the main varying are masks, namely, which functional connections form masks and which functional connections exist for each of the operands used in the execution unit. The utilization of the operands used in execution units varies and depends on the target workloads under consideration. In this case, the more often a certain operand whose read ports have been reduced is used, the more conflicts are expected for this operand. In addition, the higher the port utilization for certain operands of arithmetic units whose read ports are not reduced, the less likely we are to read from the corresponding read ports the operands requested in memory access execution units that have functional connections to this port.

| C | | 0 | | 1 | | 2 | | 3 | |
|---|---|---|---|---|---|---|---|---|---|
|   |   | 0 | 1 | 2 | 3 | 4 | 5 | 6 | 7 |
| 4 | s1 | **1** | 0 | 0 | 0 | 0 | 0 | 0 | 0 |
|   | s2 | 0 | **1** | 0 | 0 | 0 | 0 | 0 | 0 |
| 5 | s1 | 0 | 0 | **1** | 0 | 0 | 0 | 0 | 0 |
|   | s2 | 0 | 0 | 0 | **1** | 0 | 0 | 0 | 0 |
| 6 | s1 | 0 | 0 | 0 | 0 | **1** | 0 | 0 | 0 |
|   | s2 | 0 | 0 | 0 | 0 | 0 | **1** | 0 | 0 |
| 7 | s1 | 0 | 0 | 0 | 0 | 0 | 0 | **1** | 0 |
|   | s2 | 0 | 0 | 0 | 0 | 0 | 0 | 0 | **1** |
| 8 | s2 | 0 | 0 | 0 | 0 | **1** | 0 | 0 | 0 |

Figure 5. Example of the uniform symmetric read-port-count reduction scheme with 1 functional connection

| C | | 0 | | 1 | | 2 | | 3 | |
|---|---|---|---|---|---|---|---|---|---|
|   |   | 0 | 1 | 2 | 3 | 4 | 5 | 6 | 7 |
| 4 | s1 | 0 | **1** | 0 | 0 | **1** | 0 | 0 | 0 |
|   | s2 | 0 | 0 | 0 | **1** | 0 | 0 | **1** | 0 |
| 5 | s1 | 0 | 0 | **1** | 0 | 0 | 0 | 0 | **1** |
|   | s2 | **1** | 0 | 0 | 0 | 0 | **1** | 0 | 0 |
| 6 | s1 | 0 | 0 | 0 | **1** | 0 | 0 | **1** | 0 |
|   | s2 | 0 | **1** | 0 | 0 | **1** | 0 | 0 | 0 |
| 7 | s1 | **1** | 0 | 0 | 0 | 0 | **1** | 0 | 0 |
|   | s2 | 0 | 0 | **1** | 0 | 0 | 0 | 0 | **1** |
| 8 | s2 | **1** | 0 | 0 | 0 | 0 | **1** | 0 | 0 |

Figure 6. Example of the uniform symmetric read-port-count reduction scheme with 2 functional connections

In uniform symmetric schemes with a fixed number of connections, the main parameters of variation are masks, namely, which functional connections form masks and which functional connections exist for each of the operands of the execution units.

| C | | 0 | | 1 | | 2 | | 3 | |
|---|---|---|---|---|---|---|---|---|---|
|   |   | 0 | 1 | 2 | 3 | 4 | 5 | 6 | 7 |
| 4 | s1 | **1** | 0 | 0 | **1** | **1** | 0 | 0 | **1** |
|   | s2 | 0 | **1** | **1** | 0 | 0 | **1** | **1** | 0 |
| 5 | s1 | 0 | **1** | **1** | 0 | 0 | **1** | **1** | 0 |
|   | s2 | **1** | 0 | 0 | **1** | **1** | 0 | 0 | **1** |
| 6 | s1 | 0 | **1** | **1** | 0 | 0 | **1** | **1** | 0 |
|   | s2 | **1** | 0 | 0 | **1** | **1** | 0 | 0 | **1** |
| 7 | s1 | 0 | **1** | **1** | 0 | 0 | **1** | **1** | 0 |
|   | s2 | **1** | 0 | 0 | **1** | **1** | 0 | 0 | **1** |
| 8 | s2 | 0 | **1** | **1** | 0 | 0 | **1** | **1** | 0 |

Figure 7. Example of the uniform symmetric read-port-count reduction scheme with 4 functional connections

The utilization of the operands of the execution units is different and depends heavily on the target workloads. In this case, the more often a certain operand of execution units whose read ports have been reduced is used, the more conflicts are expected to be received for this operand. In addition, the higher the port utilization for certain operands of arithmetic units whose read ports are not reduced, the less likely we are to read from the corresponding read ports.

Without any applied port reduction techniques, the distribution for the operands of the arithmetic execution units by the frequency of use (utilization) of the corresponding read ports is shown in Fig. 8. The arithmetic units executing the same classes of operations are shown in the same colors. The utilizations for the operands of the units executing symmetric sets of operations are practically the same, since at the allocation stage, to increase the machine's performance, the execution control logic evenly distributes incoming operations among issue ports.

| 0 | | 1 | | 2 | | 3 | |
|---|---|---|---|---|---|---|---|
| 0 | 1 | 2 | 3 | 4 | 5 | 6 | 7 |
| 13.4% | 14.8% | 9.9% | 11.9% | 13.4% | 14.8% | 9.9% | 11.9% |

Figure 8. Utilization of the PRF read ports corresponding to different operands in arithmetic instructions

As we can see, the read ports corresponding to the $0^{th}$ and $2^{nd}$ issue ports are the most loaded: the utilization of the read port for the first operand is 13.4%, and the utilization of the read port by the second operand is 14.8%. At the same time, the read ports in arithmetic operations are more often used for the second operands, since the first operands more often take the register values from the data transfer system between processor stages.

According to the conducted analysis, the distribution of memory access operands by the frequency of read accesses to the integer physical register file is shown in Fig. 9.

| 4 | | 5 | | 6 | | 7 | | 8 |
|---|---|---|---|---|---|---|---|---|
| s1 | s2 | s1 | s2 | s1 | s2 | s1 | s2 | s2 |
| 21.0% | 8.1% | 21.1% | 8.1% | 11.6% | 5.3% | 11.6% | 5.3% | 7.8% |

Figure 9. Utilization of the PRF read ports corresponding to different operands in memory access instructions

110

As we can see from Fig. 9, the operands corresponding to the read ports with the highest load are the base operands of load instructions with a utilization of 21.0%. The operands corresponding to the read ports with the lowest utilization are the index operands of STA instructions with a utilization of 5.3%.

When selecting a specific mask for the port reduction scheme, the probability that at least one of the read ports featured in the functional connections in this mask will be free is determined by the utilization distribution for the arithmetic operands shown in Fig. 8. In a simple case, the occupancy of a specific mask, determined by the probability of getting a conflict when reading a register value using this mask, can be found as the sum of the occupancies for the incoming functional connections with the corresponding operands of the arithmetic units. Thus, the masks of the read-port-count reduction schemes vary in their expected occupancy. Using the arithmetic operand utilization distribution table, it is necessary to select masks with the lowest occupancy to mitigate the introduced performance degradation as much as possible when using the read port reduction scheme. At the same time, for a given number of functional connections, the masks with uniform expected occupancies should be selected to prevent a large drop in performance on specialized workloads. For the same reason, it is necessary to select functional connections with operands belonging to units executing operations from different classes. This will mitigate a large performance degradation on workloads in which the input instruction program flow is uniform in terms of the classes of operations present (for example, most arithmetic operations are shift operations). Based on the proposed set of rules minimizing the probability of getting a conflict for a certain mask, the set of masks for the efficient uniform symmetric scheme with 2 functional connections will be the following: (0, 3), (4, 7), (1, 2), (5, 6).

even though the masks obtained for the uniform symmetric schemes using these rules have the smallest variance of the expected load, for schemes with a large number of connections (4 functional connections in a mask) or with a sufficiently small number of connections (1 functional connection in masks), the pairwise difference in the values of the expected loads between different masks will be significant. To mitigate performance degradation when using read-port-count reduction schemes, it is necessary to assign masks with the lowest expected load to the most used operands of memory access execution units. In this case, 2 masks with an expected load of 24.7% should be assigned to the base operands of load instructions. Furthermore, different masks should be assigned to the same operands. For schemes with a small total number of masks, the set of operands corresponding to the same mask should be as non-uniform as possible in terms of operand types and memory access execution units. These ideas correspond to the ones introduced in [16].

However, simply mapping a set of selected masks sorted in the increasing order of the expected utilization onto a set of memory access instruction operands sorted in the decreasing order of utilization (frequency of access to the physical register file) is insufficient to obtain efficient schemes that mitigate performance degradation. The same selected mask should be mapped onto the memory access operands in groups organized in such a way that, first, the average expected mask load, calculated taking into account the utilization of the corresponding operands, is minimal, and, second, the variance of the expected utilization in memory access operands is minimal. This allows balancing the read-port-count reduction schemes so that frequently used memory access operands are not assigned to the same masks, and, hence, the mathematical expectation of the number of conflicts for read ports is decreased. For example, for the scheme with 2 functional connections, the groups consist of 2 and 3 operands and are as follows:

1. The base operand of load instructions and the index operand of STA instructions

2. The index operand of load instructions and the base operand of STA instructions

3. The index operand of load instructions, the base operand of STA instructions, and the operand of STD instructions.

Thus, the complete heuristic algorithm for constructing efficient uniform symmetric read-port-count reduction schemes consists of the following stages:

1. Obtain the distribution of read port utilization for the operands of arithmetic instructions and operands of memory access instructions for the target workloads.

2. According to the proposed rules, determine the set of scheme masks for a given number of functional connections

3. Based on the proposed distribution of operands by groups, match the found masks to the operands of memory access instructions, minimizing the mathematical expectation of the number of conflicts for read ports.

Using the proposed heuristic algorithm, uniform symmetric read-port-count reduction schemes were constructed for the number of functional connections 1, 2, and 4. The constructed schemes are shown in Fig. 10, Fig. 11, and Fig. 12, respectively.

| C | | 0 | | 1 | | 2 | | 3 | |
|---|---|---|---|---|---|---|---|---|---|
| | | 0 | 1 | 2 | 3 | 4 | 5 | 6 | 7 |
| 4 | s1 | 0 | 0 | 1 | 0 | 0 | 0 | 0 | 0 |
| | s2 | 1 | 0 | 0 | 0 | 0 | 0 | 0 | 0 |
| 5 | s1 | 0 | 0 | 0 | 0 | 0 | 0 | 1 | 0 |
| | s2 | 0 | 0 | 0 | 0 | 1 | 0 | 0 | 0 |
| 6 | s1 | 0 | 0 | 0 | 1 | 0 | 0 | 0 | 0 |
| | s2 | 0 | 1 | 0 | 0 | 0 | 0 | 0 | 0 |
| 7 | s1 | 0 | 0 | 0 | 0 | 0 | 0 | 0 | 1 |
| | s2 | 0 | 0 | 0 | 0 | 0 | 1 | 0 | 0 |
| 8 | s2 | 0 | 1 | 0 | 0 | 0 | 0 | 0 | 0 |

Figure 10. Efficient uniform symmetric read-port-count reduction scheme with 1 functional connection

As we stated earlier, uniform symmetric read-port-count reduction schemes with the same number of functional connections have the same estimated complexity. The estimates of the critical path length for the constructed schemes shown in Fig. 10, Fig. 11, and Fig. 12 are 2, 5, and 9 time units, respectively.

| C | | 0 | | 1 | | 2 | | 3 | |
|---|---|---|---|---|---|---|---|---|---|
| | | 0 | 1 | 2 | 3 | 4 | 5 | 6 | 7 |
| 4 | s1 | 0 | 1 | 1 | 0 | 0 | 0 | 0 | 0 |
| | s2 | 0 | 0 | 0 | 0 | 1 | 0 | 0 | 1 |
| 5 | s1 | 0 | 0 | 0 | 0 | 0 | 1 | 1 | 0 |
| | s2 | 1 | 0 | 0 | 1 | 0 | 0 | 0 | 0 |
| 6 | s1 | 0 | 0 | 0 | 0 | 1 | 0 | 0 | 1 |
| | s2 | 0 | 1 | 1 | 0 | 0 | 0 | 0 | 0 |
| 7 | s1 | 1 | 0 | 0 | 1 | 0 | 0 | 0 | 0 |
| | s2 | 0 | 0 | 0 | 0 | 0 | 1 | 1 | 0 |
| 8 | s2 | 1 | 0 | 0 | 1 | 0 | 0 | 0 | 0 |

Figure 11. Efficient uniform symmetric read-port-count reduction scheme with 2 functional connections

| C | | 0 | | 1 | | 2 | | 3 | |
|---|---|---|---|---|---|---|---|---|---|
| | | 0 | 1 | 2 | 3 | 4 | 5 | 6 | 7 |
| 4 | s1 | 0 | 1 | 1 | 0 | 1 | 0 | 0 | 1 |
| | s2 | 1 | 0 | 0 | 1 | 0 | 1 | 1 | 0 |
| 5 | s1 | 1 | 0 | 0 | 1 | 0 | 1 | 1 | 0 |
| | s2 | 0 | 1 | 1 | 0 | 1 | 0 | 0 | 1 |
| 6 | s1 | 0 | 1 | 1 | 0 | 1 | 0 | 0 | 1 |
| | s2 | 1 | 0 | 0 | 1 | 0 | 1 | 1 | 0 |
| 7 | s1 | 1 | 0 | 0 | 1 | 0 | 1 | 1 | 0 |
| | s2 | 0 | 1 | 1 | 0 | 1 | 0 | 0 | 1 |
| 8 | s2 | 1 | 0 | 0 | 1 | 0 | 1 | 1 | 0 |

Figure 12. Efficient uniform symmetric read-port-count reduction scheme with 4 functional connections

Performance evaluations for the three constructed read-port-count reduction schemes are presented in Fig. 13. The scheme with 1 functional connection results in -1.6% geomean IPC performance degradation on SPECrate CPU 2017 Integer workloads, while the schemes with 2 and 4 functional connections allow reducing the number of read ports to the integer PRF by a factor of two with the respective geomean performance degradations of -0.2% IPC and -0.1% IPC on SPECrate CPU 2017 Integer workloads.

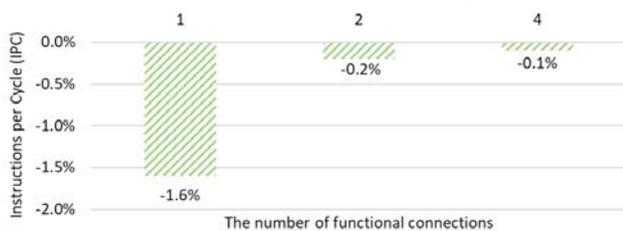

Figure 13. Geomean performance degradations with the constructed uniform symmetric read-port-count reduction schemes for the integer PRF (17 → 8 read ports)

The worst-case performance degradations on individual workloads for the constructed schemes with 2 and 4 functional connections are -14% IPC and -5% IPC, respectively.

## VII. CONCLUSION

The number of ports to the physical register file in superscalar out-of-order processors significantly impacts the power consumption and the required chip area. In this work, we introduce a practical approach to constructing low-complexity read-port-count reduction schemes for the integer physical register in superscalar CPUs with single-thread out-of-order execution.

The constructed read-port-count reduction schemes show geomean performance degradations as low as -0.1% IPC across SPECrate CPU 2017 Integer workloads with the number of read ports reduced from 17 to 8.

REFERENCES

[1] R. Shioya, K. Horio, M. Goshima, and S. Sakai, "Register cache system not for latency reduction purpose," in *2010 43rd Annual IEEE/ACM International Symposium on Microarchitecture*, Atlanta, GA, USA, 2010, pp. 301-312, DOI: 10.1109/MICRO.2010.43.

[2] S. Rixner, W. J. Dally, B. Khailany, P. Mattson, U. J. Kapasi and J. D. Owens, "Register organization for media processing," in *Proc. of the Sixth International Symposium on High-Performance Computer Architecture*, Touluse, France, 2000, pp. 375-386, DOI: 10.1109/HPCA.2000.824366.

[3] S. Thoziyoor, N. Muralimanohar, J. Ahn, and N. Jouppi, "Cacti 5.1," HP Laboratories, Palo Alto, Tech. Rep. HPL-2008-20, 2008.

[4] S. Mittal, "A Survey of Techniques for Designing and Managing CPU Register File," *Concurrency and Computation: Practice and Experience*, vol. 29, no. 4, pp. 1-23, 2017, DOI: 10.1002/cpe.3906.

[5] S. Sirsi and A. Aggarwal, "Exploring the limits of port reduction in centralized register files," in *2009 22nd International Conference on VLSI Design*, New Delhi, India, 2009, pp. 535-540, DOI: 10.1109/VLSI.Design.2009.29.

[6] R. Balasubramonian, S. Dwarkadas, and D. H. Albonesi, "Reducing the complexity of the register file in dynamic superscalar processors," in *Proc. of the 34th ACM/IEEE International Symposium on Microarchitecture*, Austin, TX, USA, 2001, pp. 237-248, DOI: 10.1109/MICRO.2001.991122.

[7] J. A. Swensen and Y. N. Patt, "Hierarchical registers for scientific computers," in *Proc. of the 2nd International Conference on Supercomputing*, St. Malo, France, 1988, pp. 346-354, DOI: 10.1145/55364.55398.

[8] J.-L. Cruz, A. Gonzalez, M. Valero, and N. P. Topham, "Multiple-banked register file architectures," in *Proc. of 27th International Symposium on Computer Architecture*, Vancouver, BC, Canada, 2000, pp. 316-325.

[9] R. Nalluri, R. Garg, and P. R. Panda, "Customization of Register File Banking Architecture for Low Power," in *20th International Conference on VLSI Design held jointly with 6th International Conference on Embedded Systems*, Bangalore, India, 2007, pp. 239-244, doi: 10.1109/VLSID.2007.58.

[10] S. Wang, H. Yang, J. Hu, and S. G. Ziavras, "Asymmetrically Banked Value-Aware Register Files," in *IEEE Computer Society Annual Symposium on VLSI*, Porto Alegre, Brazil, 2007, pp. 363-368, DOI: 10.1109/ISVLSI.2007.27.

[11] R. Sangireddy and A. K. Somani, "Exploiting quiescent states in register lifetime," in *Proc. of the IEEE International Conference on Computer Design: VLSI in Computers and Processors*, San Jose, CA, USA, 2004, pp. 368-374, DOI: 10.1109/ICCD.2004.1347948.

[12] A. Aggarwal and M. Franklin, "Energy efficient asymmetrically ported register files," in *Proc. of the 21st International Conference on Computer Design*, San Jose, CA, USA, 2003, pp. 2-7, DOI: 10.1109/ICCD.2003.1240865.

[13] T. M. Jones, M. F. O'Boyle, J. Abella, A. Gonzalez, and O. Ergin, "Energy-efficient register caching with compiler assistance," *ACM Trans. Archit. Code Optim*, vol. 6, no. 4, Article 13, 2009, DOI: 10.1145/1596510.1596511.

[14] J. A. Butts and G. S. Sohi, "Use-based register caching with decoupled indexing," in *Proc. of the 31st Annual International Symposium on Computer Architecture,* Munich, Germany, 2004, pp. 302-313, DOI: 10.1109/ISCA.2004.1310783.

[15] D. A. Los and I.V. Smirnov, "Caching physical register file in a modern superscalar microprocessor," (in Russian), in *Proc. of the 61th MIPT Scientific Conference. Radio engineering and computer technologies*, Moscow, Russia, 2018, pp. 18-19.

[16] N. Goel, A. Kumar, and P. R. Panda, "Shared-port register file architecture for low-energy VLIW processors," *ACM Trans. Archit. Code Optim*, vol. 11, no. 1, Article 1, 2014, DOI: 10.1145/2533397.

[17] J. Busek *et al.,* "SPEC CPU2017: Next-Generation Compute Benchmark," in *Companion of the 2018 ACM/SPEC International Conference on Performance Engineering*, Berlin, Germany, 2018, pp. 41-42, DOI: 10.1145/3185768.3185771.

[18] V.P. Nelson *et al.,* "Digital Circuit Analysis and Design," Prentice Hall, 1995, p. 234